\documentclass[a4paper]{jpconf}
\usepackage{graphicx}

\begin{document}
\title{Entanglement engineering and topological protection by discrete-time quantum walks}

\author{Simon Moulieras$^{1}$, Maciej Lewenstein$^{1,2}$, Graciana Puentes$^{1}$}

\address{$^{1}$ ICFO - Institut de Ciencies Fotoniques, Mediterranean Technology Park, 08860 Castelldefels, Barcelona, Spain\\
\noindent $^{2}$ ICREA - Institucio Catalana de Recerca i Estudis Avancats, 08015 Barcelona, Spain}

\ead{graciana.puentes@icfo.es}





\begin{abstract}
Discrete-time quantum walks (QWs) represent robust and versatile platforms for the controlled engineering of single particle quantum dynamics, and have attracted special attention due to their algorithmic applications in quantum information science. Even in their simplest 1D architectures, they display complex topological phenomena, which can be employed in the systematic study of topological quantum phase transitions \cite{Kitagawa}.~Due to the exponential scaling in the number of resources required, most experimental realizations of QWs up to date have been limited to single particles, with only a few implementations involving correlated quantum pairs. In this article we study applications of quantum walks in the controlled dynamical engineering of entanglement in bipartite bosonic systems. We show that quantum walks can be employed in the transition from mode entanglement, where indistinguishability of the quantum particles plays a key role, to the standard type of entanglement associated with distinguishable particles. We also show that, by carefully tailoring the steps in the QWs, as well as the initial state for the quantum walker, it is possible to preserve the entanglement content by topological protection. The underlying mechanism that allows for the possibility of both entanglement engineering and entanglement protection is the strong ``spin-orbit" coupling induced by the QW. We anticipate that the results reported here can be employed for the controlled emulation of quantum correlations in topological phases. 
\end{abstract}

\section{Introduction}

Discrete-time quantum walks (QWs), originally inspired by the Feynman path-integral approach for computing observables \cite{Feynman}, are the quantum extension of classical random walks, where classical probabilities are replaced by quantum imaginary amplitudes, so that the multiple paths of the quantum walker can interfere, a feature which results in the probability of the quantum walker being quite distinct from their classical counterpart \cite{Aharonov}.~Within the realm of classical statistical physics, random walks have been extensively employed as fundamental models describing the complex dynamics of several systems of interest, ranging from transport properties in biological and material compounds, to the dynamical evolution of financial markets. On the other hand, the analogous quantum walks have been used to model a large number of dynamical physical processes involving quantum coherence, such as photosynthesis \cite{photonsyn}, diffusion \cite{diffusion}, and vortex transport \cite{vortex}.  In recent years, QWs have attracted significant attention as they provide a universal platform for the study of quantum information protocols, allowing for speed-up in search algorithms \cite{shenvi, potocek}, and for universal simulation of quantum circuits \cite{childs}. Moreover, it has recently become apparent that QWs can be employed for the efficient dynamical exploration of a wide range of non-trivial topological phases in 1D and 2D \cite{Kitagawa, obuse}. In particular, the simplest 1D random walk that presents non-trivial topological features is analogous to the SSH model in polyacetylene \cite{SSH}. Several experimental realizations of QWs have readily been accomplished, employing trapped ions \cite{ions}, ultra-cold atoms in optical lattices \cite{coldatoms}, and photons \cite{photons}. The latter are of particular interest, as light particles can be easily produced, manipulated and detected. However, due to the exponential growth in the Hilbert space size and in the number resources, only a few QW implementations up to date involve correlated quantum particles \cite{Science, omar}.\\

 In this paper, we analyze the use of QWs in the controlled engineering of entanglement between bosonic quantum particles. We discuss in detail the different types of entanglement that can be observed between  the quantum particles, namely mode entanglement of indistinguishable particles, and standard entanglement between separate distinct particles, the latter being a  fundamental resource for quantum communications. The entanglement of elementary states of indistinguishable particles was discussed in detail in the classic book by A. Peres \cite{peres}, and has been the subject of  continued active research \cite{schliemann, zanardi}. Here we show that quantum walks can be employed to engineer controlled transitions between different types of entanglement. This can be explained in terms of coupling between different internal degrees of freedom in the quantum walker, such as spin and spatial or orbital degrees of freedom, a result which can be exploited for the controlled dynamical engineering of entanglement. This form of ``spin-orbit" coupling mechanism induced by the quantum walk is also at the heart of topological phenomena, as first introduced in the seminal work by Kitagawa and co-workers \cite{Kitagawa}. \\
 
 Here we demonstrate that by carefully engineering the parameters in the quantum walk, such as the quantum coin, and the initial state of the quantum walker, it is possible to preserve the entanglement content from noise and perturbations, a feature  which can, in turn, be explained in terms of topological protection. The term ``topological protection", usually refers to immunity to local noise and back-scattering of highly symmetrical ground state wave-functions of quantum systems characterized by some form of topological order. Usually, such topological bound states display strong spin-orbit coupling and are protected against perturbations by time-reversal symmetry. The idea of entanglement preservation by topological protection, is to prepare the initial state of the system in such a way that there is a significant overlap with the topological bound states. Here, we show that by tailoring the ``band-structure" characterizing the QW, this can be done by choosing particular values for the quantum coin, it is possible to simulate effective Hamiltonians which are gapless, showing a Dirac-like dispersion relation. In the vicinity of such Dirac points the emergence of robust localized bound states is expected. Therefore, by encoding the initial state of the bipartite quantum walker in the topologically protected bound state basis, it possible to protect its entanglement content.\\ 
 
The article is structured as follows, in Section II we review the formalism for quantum walks of single and composite systems, as well as the relevant entanglement measures. Next in Section III, we present our results for entanglement engineering in systems of  indistinguishable particles. In Section IV, we present application of quantum walks in topological protection. Finally, in Section V we present the conclusions. 

\bigskip


\section{Formalism}

\subsection{Basic Formalism: QW for single 2-level systems in 1D}

The simplest quantum walk consists of a protocol for a 2-level particle ($\uparrow$,$\downarrow$), moving in dimension. The unitary evolution operator $U$ for a single step in the quantum walk is given by:

\begin{equation}
\label{eq:1}
\hat{U}=\hat{T}\hat{R}_{\vec{v}}(\theta),
\end{equation}

where $\hat{T} = \sum_x |\uparrow \rangle \langle \uparrow|  \otimes  |x + 1 \rangle \langle x| + |\downarrow \rangle \langle \downarrow |  \otimes  |x-1 \rangle \langle x |$, is a spin-dependent translation by one site $x$, $\hat{R}_{\vec{v}}(\theta)=\exp{\left(-i \theta \vec{v} \cdot \hat{\vec{\sigma}}/2 \right)}$ is the coin operator given by an arbitrary SU(2) rotation around an unitary vector $\vec{v}$ in the Bloch sphere, and $\hat{\vec{\sigma}}=(\hat{\sigma}_{x},\hat{\sigma}_{y},\hat{\sigma}_{z})$ are the Pauli matrices. It is the quantum equivalent of randomly tossing a coin and choosing which way the particle will move. Conventionally, the rotation is set to $\hat{R}_{y}=e^{-i \theta \hat{\sigma}_{y}/2}$.  The unitary evolution operator $\hat{U}$ can be considered as a stroboscopic regularization of a static effective Hamiltonian $\hat{H}_{\mathrm{eff}}$ given by:

\begin{equation}
\hat{U}=e^{-i \hat{H}_{\mathrm{eff}}}.
\end{equation}


We note that the spin-dependent translation can be written in quasi-momentum space as $\hat{T}=\int_{-\pi} ^{\pi} dk e^{i k \hat{\sigma}_{z}} \otimes |k \rangle \langle k|$, which readily reveals the spin-orbit coupling mechanism which is at the heart of the entanglement engineering and topological protection mechanisms described in this paper. In quasi-momentum space, the effective Hamiltonian up to a constant energy offset, can be written as:

\begin{equation}
\hat{H}_{\mathrm{eff}}=\int_{-\pi} ^{\pi} dk[E(k) \vec{n}(k) \cdot \hat{\vec{\sigma}}] \otimes |k\rangle \langle k|.
\end{equation}

Now $\vec{n}(k)$ defines the quantization axis for the spinor eigenstates at each quasi-momentum $k$. For a coin operator different from the identity operator, the quasi-energy is given by $\cos(E(k))=\cos(\theta/2)\cos(k)$, typically corresponding to a 2-band structure, for a 2-level system. The quasi-energy spectrum is plotted in Figure 1 for different values of the coin parameter $\theta$. For $|\theta| >0$ the spectrum has a gap, and for $\theta \approx 0, 2 \pi$ the spectrum becomes gapless, and presents a Dirac-like dispersion relation, where the quasi-energy vs quasi-momentum are linearly related. This already explains why QWs in their simplest architectures can be used to efficiently simulate dynamical properties of complex condensed matter system with topological properties. \\

\begin{figure} [h!]
\label{fig:5}
\begin{center}
\includegraphics[width=0.5\linewidth]{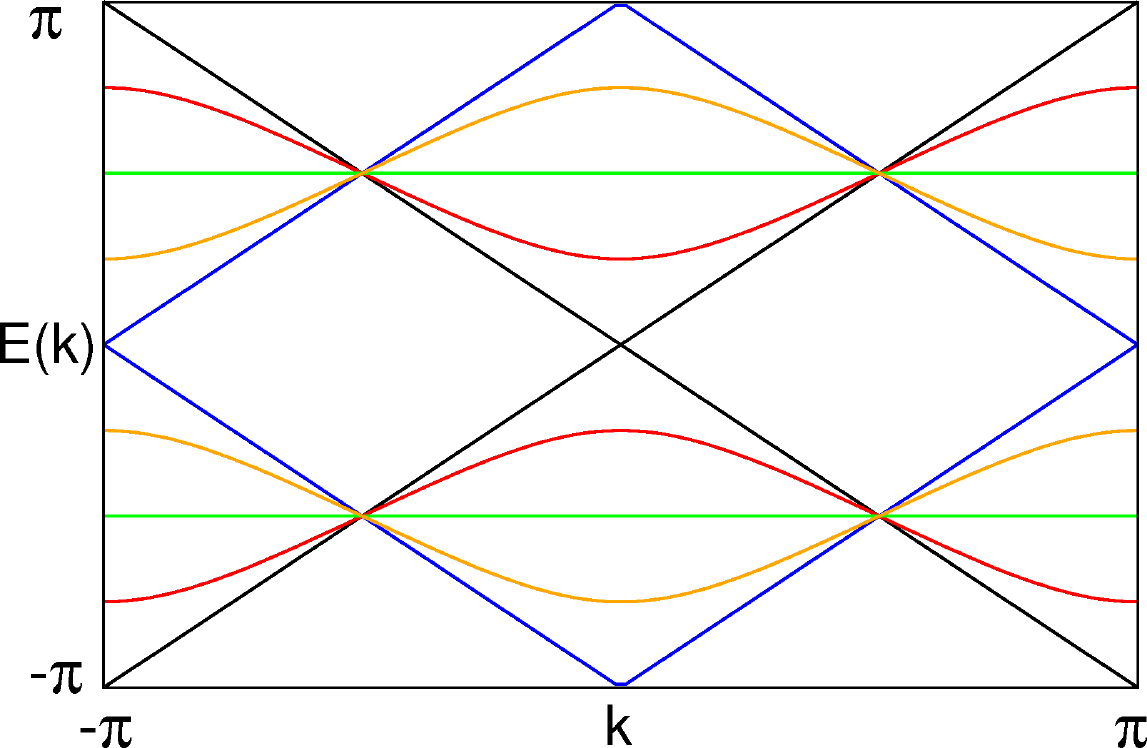}
\caption{Band structures for the simplest 1D QW in a 2-level system. By tuning the quantum coin parameter $\theta$ it is possible simulate a gapped system, or a gapless system with Dirac points. $\theta=0$ (black), $\theta=\pi/2$ (red), $\theta=\pi$ (green), $\theta=3\pi/2$ (orange), $\theta=2\pi$ (blue).}
\end{center}
\end{figure}

In general,  non-trivial topological character requires at least two independent parameters, so that an adiabatic variation of these parameters leads to the accumulation of a Berry phase \cite{berry}. This can be done by introducing a second rotation $R_{y}(\theta_{2})$, which is called in the literature the ``split-step'' quantum walk \cite{Kitagawa}.


\subsection{Split-step quantum walk}

The split-step QW protocol reads :

\begin{equation}
\hat{U}(\theta_{1}, \theta_{2})=\hat{T}_{\downarrow}\hat{R}_{y}(\theta_{2})\hat{T}_{\uparrow}\hat{R}_{y}(\theta_{1}).
\end{equation}

where:

\begin{eqnarray}
\hat{T_{\uparrow}} &=& \sum_x |\uparrow \rangle \langle \uparrow|  \otimes  |x + 1 \rangle \langle x| + |\downarrow \rangle \langle \downarrow|  \otimes  |x \rangle \langle x| ,\\
\hat{T_{\downarrow}} &=& \sum_x |\downarrow \rangle \langle \downarrow|  \otimes  |x - 1 \rangle \langle x| + |\uparrow \rangle \langle \uparrow|  \otimes  |x \rangle \langle x| .
\end{eqnarray}

It has been shown analytically \cite{Kitagawa} that this system still presents chiral symmetry, meaning that the positions $\{ \vec{n}(k) \}$ of the eigenstates of $\hat{U}(\theta_1,\theta_2)$ lye on a circle of the Bloch sphere. Nevertheless, depending on the values of $\theta_1$ and $\theta_2$, the set $\{ \vec{n}(k) \}$ can describe the full circle ($W=1$), or not ($W=0$), where $W$ is the winding number associated to the topological structure of the spectrum. The spectrum presents a gap in both cases $|\tan(\theta_2/2)/\tan(\theta_1/2)|<1$ ($W=1$) and  $|\tan(\theta_2/2)/\tan(\theta_1/2)|>1$ ($W=0$). When $|\tan(\theta_2/2)/\tan(\theta_1/2)|=1$, the gap closes (at quasi-energy $0$ or $\pi$), and one can not properly define a winding number. In this case, the corresponding $0$(resp. $\pi$) quasi-energy state corresponds to a so called topological bound state. The topological phase diagram with two coin parameters becomes a mesh-grid of alternating squares, characterized by alternating topological numbers $W=0,1$ (see fig. (2)). We can notice that when $\theta_2=0$, we recover the standard quantum walk protocol operator.\\

\begin{figure} [h!]
\begin{center}
\includegraphics[width=0.5\linewidth]{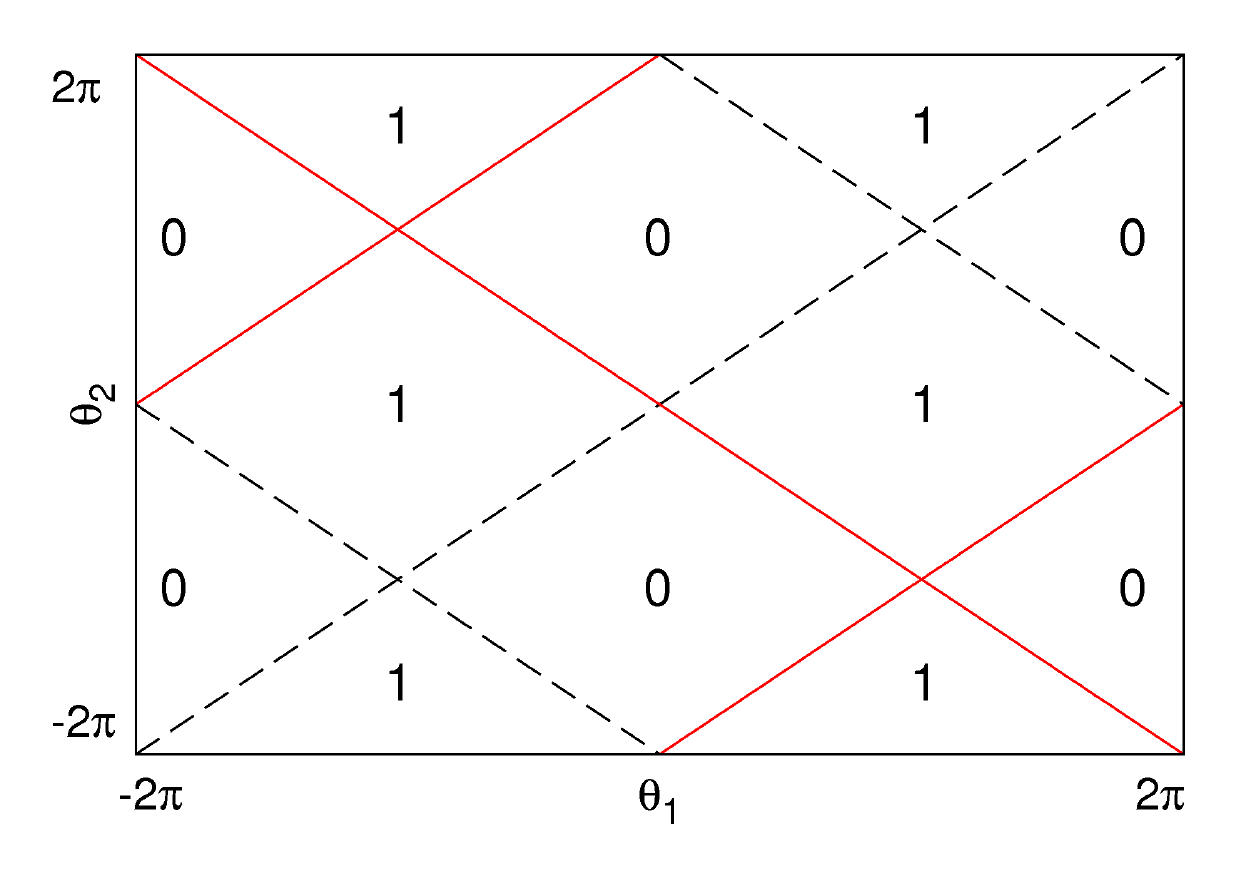}
\caption{Diagram representing the topological nature of the split-step QW. Full red (resp. dashed black) lines represent configurations that give a gapless spectrum where the gap closes at quasi-energy 0 (resp. $\pi$). The numbers (0 and 1) written in the different zone are the winding numbers of the given choice of angles ($\theta_1$, $\theta_2$).}
\end{center}
\end{figure}

In order to experimentally observe the presence of the topological nature of the split-step QW, one can decide to make one of the rotations inhomogeneous; for instance, make $\theta_2$ vary as a function of $x$, going from $\theta_2^-$ (when $x \to -\infty $) to $\theta_2^+$ (when $x \to +\infty $). If one line where the gap closes is crossed by the trajectory of ($\theta_1$, $\theta_2$), we know that the local spectra are gapped far from the crossing point, and gapless in the vicinity of this point (see fig. (3)). The associated bound state (say of $0$ quasi-energy) will be localized around the zone where the ($\theta_1$, $\theta_2$) trajectory crosses the line (say $x=0$). Already at a few sites from $x=0$, there is no arbitrary small quasi-energy states that could permit the propagation of the $0$ quasi-energy component of the wavepacket, and consequently the overlap between the initial wavepacket and the localized bound state is topologically ``trapped'' around $x=0$. Then, after propagation, the probability to detect the particle localized around $x=0$ remains significant compared to the case in which bound states are not present in the spectrum. Indeed, the basic feature of the QW evolution of an initially localized (around $x=0$) particle is a ballistic expansion which is in general anisotropic (not symmetric in 1D) and where the probability for the particle to stay at its initial position decreases very quickly. 
This difference of behaviors has been indeed experimentally observed \cite{KitagawaNatComm}, and clearly shows the important impact that the topological structure of the system has on the dynamics. Note that the trajectory ($\theta_1$, $\theta_2$) has to keep $\theta_1$ uniform in order to preserve the chiral symmetry of the system, but the variations of ($\theta_2$) can be arbitrary. \\

\begin{figure} [h!]
\begin{center}
\includegraphics[width=0.5\linewidth]{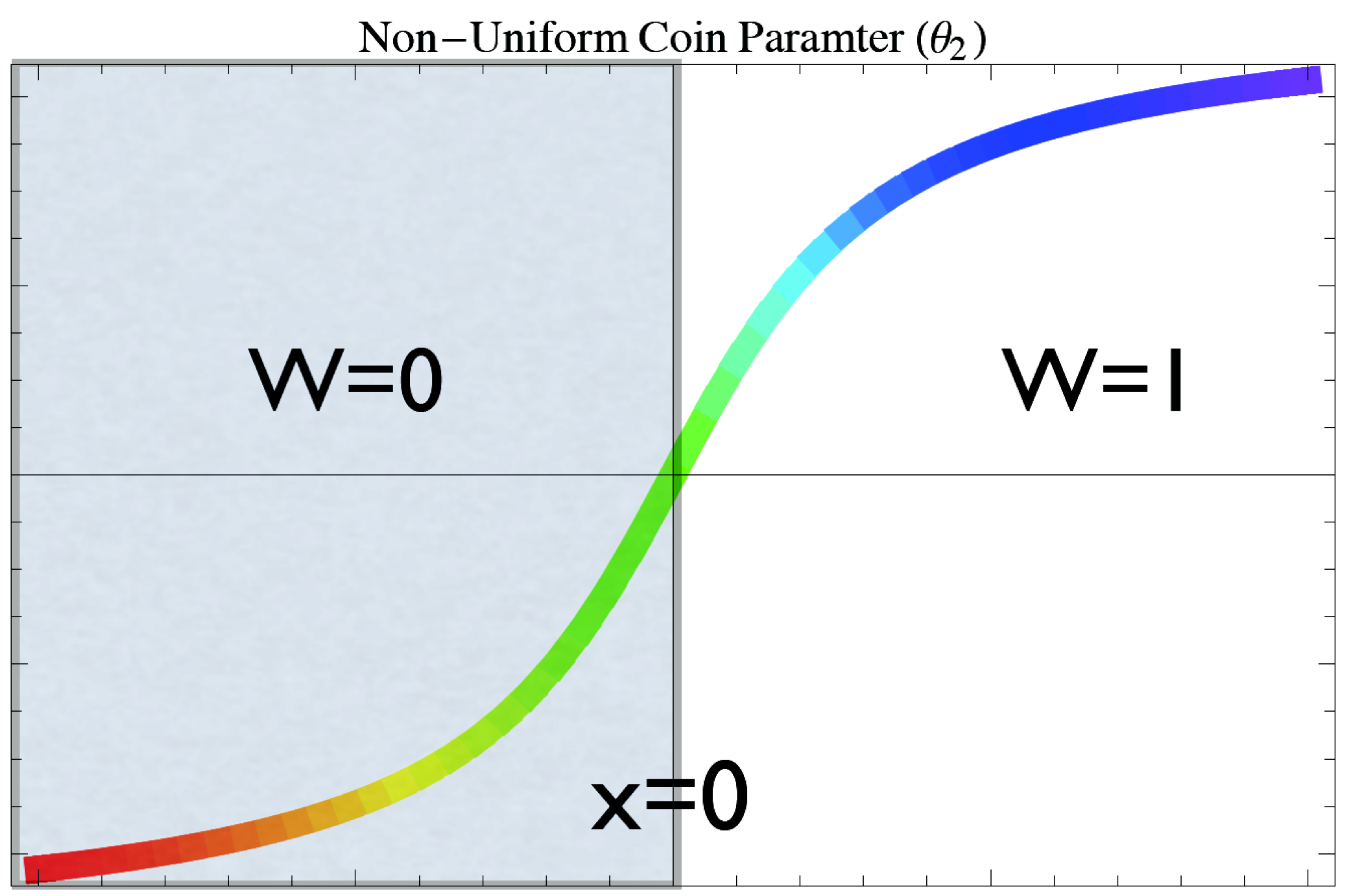}
\caption{Spatially varying coin parameter $\theta_{2}(x)$ characterizing the inhomogenous split-step quantum walk. The horizontal axis represent the spatial degrees of freedom for the 1D QW ($x$-direction).~This inhomogeneous step allows to assign different values to the parameter $\theta_{2}^{-,+}$ across topological sectors characterized by different winding numbers ($W=0,1$) and in this way create a boundary. At the spatial location of the topological boundary $(x=0)$ a topologically protected bound state is expected. This feature can be used for entanglement topological protection, by encoding the initial state of the quantum walker in a  localized state in the vicinity of this boundary.}
\end{center}
\end{figure}

\subsection{Basic Formalism: QW for two identical 2-level systems in 1D}

We consider now the dynamics of two identical bosonic particles, such as linearly polarized photons, propagating in the same QW protocol, represented by identical unitary evolution operators $\hat{U}(\theta_1,\theta_2)$, as defined in the previous subsection. Since photons do not interact, we focus on the evolution of entanglement on the two-particle dynamics in order to analyze to impact of quantum correlations in the system. \\

The Hilbert space of a bipartite system is given by $\mathcal{H}=\mathcal{H}_{A} \otimes \mathcal{H}_{B}$, and the step operator is now $\hat{U}=\hat{U}_{A}\otimes \hat{U}_{B}$. In the case of indistinguishable particles, as considered in this work, the structure of the Hilbert space is different because both particles ``live'' in the same Fock space. The adapted formalism in this context is the second quantization, that automatically symmetrizes (resp. antisymmetrizes) the bosonic (resp. fermionic) wavefunctions. \\

Let us first consider the case that the two photons are at the same spatial location $x$, can be described by identical frequencies (i.e., monochromatic case), and can only be labeled by the polarization degree of freedom (modes $\uparrow$ and $\downarrow$). Any bosonic two-particle wavefunction can be built by the superposition of the states of the canonical symmetric basis composed by 
$\hat{a}^\dag_\uparrow \hat{a}^\dag_\uparrow |mathrm{vac} \rangle = |\uparrow \uparrow \rangle$ , $\hat{a}^\dag_\downarrow \hat{a}^\dag_\downarrow |\mathrm{vac}\rangle = |\downarrow \downarrow \rangle$ and $\hat{a}^\dag_\uparrow \hat{a}^\dag_\downarrow |\mathrm{vac}\rangle = \frac{1}{\sqrt{2}} \left(|\downarrow \uparrow \rangle +|\uparrow \downarrow \rangle  \right)$ where $|\mathrm{vac}\rangle$ represents the vacuum state of the Fock space. These states can be written in a more synthetic form, in terms of mode ($\uparrow$ or $\downarrow$) with an occupation number : $\hat{a}^\dag_\uparrow \hat{a}^\dag_\uparrow |\mathrm{vac}\rangle \equiv |2 \uparrow, 0 \downarrow\rangle$, $\hat{a}^\dag_\downarrow \hat{a}^\dag_\downarrow |\mathrm{vac}\rangle \equiv |0 \uparrow, 2 \downarrow\rangle$, $\hat{a}^\dag_\uparrow \hat{a}^\dag_\downarrow |\mathrm{vac}\rangle \equiv |1 \uparrow, 1 \downarrow \rangle$. Additionally, another advantage of this notation is that it clearly appears now that the entanglement in polarization of $\frac{1}{\sqrt{2}} \left(|\downarrow \uparrow \rangle +|\uparrow \downarrow \rangle  \right)$ is just the consequence of the bosonic character of the photon pair because in the second quantization language, it is written as a product state in modes $|1 \uparrow, 1 \downarrow \rangle$, as well as the two other states mentioned above. In particular, the state $\frac{1}{\sqrt{2}} \left(|\uparrow \uparrow \rangle +|\downarrow \downarrow \rangle \right)$ which is maximally entangled in the polarization representation can be rewritten $\frac{1}{\sqrt{2}} \left(|2 \uparrow,0 \downarrow \rangle + |0 \uparrow, 2 \downarrow \rangle  \right)$ in the mode representation, which shows that this state is maximally entangled in both mode and polarization representations \cite{schliemann, zanardi}.\\ 

In order to respect the global symmetry of exchange of bosonic particles, the states describing the case in which one particle is at position $x_1$ and the other at $x_2$ are: on one hand, any possible tensorial product of the symmetric state in positions ($\frac{1}{\sqrt{2}} \left(|x_1,x_2\rangle + |x_2,x_1\rangle \right)$) and one of the symmetric state in spins ($|\uparrow \uparrow\rangle$, $|\downarrow \downarrow\rangle$, or $\frac{1}{\sqrt{2}} \left(|\downarrow \uparrow \rangle +|\uparrow \downarrow \rangle  \right)$) or, on the other hand, the tensorial product of the antisymmetric state in positions ($\frac{1}{\sqrt{2}} \left(|x_1,x_2 \rangle -|x_2,x_1\rangle  \right)$) and the antisymmetric state in spins ($\frac{1}{\sqrt{2}} \left(|\downarrow \uparrow \rangle -|\uparrow \downarrow \rangle  \right)$). Since each photon evolves with the same unitary evolution operator, an initially symmetric state with respect to the exchange of particles will remain symmetric after an arbitrary evolution time. We note that in the case of photon pairs produced by Type-II Spontaneous Parametric Down Conversion (SPDC), the two-photon state has the general form $|\psi \rangle=\int dq_{1} \int dq_{2} \int dw_{1} \int dw_{2} [\phi_{HV}(q_{1},w_{1};q_{2} w_{2})|H_{1},q_{1},w_{1};V_{2},q_{2},w_{2}\rangle +\phi_{VH}(q_{1},w_{1};q_{2},w_{2})|V_{1},q_{1},w_{1};H_{2},q_{2},w_{2} \rangle]$, where $(H,V)$ stand for orthogonal linear polarizations ($\uparrow, \downarrow$), and $(q,w)$ represent the spatial and frequency degrees of freedom, respectively.~For the case of indistinguishable particles, (i.e., no temporal or spatial walk off), the two-photon amplitude wave-functions should be identical $\phi_{VH}(q_{1},w_{1};q_{2},w_{2})\equiv \phi_{HV}(q_{1},w_{1};q_{2},w_{2})$, which forces the spatial and frequency part of the wave-function to be symmetric.~In this case, the state is maximally entangled in polarization (i.e., Bell state).~Any uncontrolled spatial or temporal walk off reduces the indistinguishability of the photon-pairs, and therefore the entanglement content.\\

In the following, in order to quantify the entanglement of a given two-particle state, we will compute the logarithmic negativity either in mode representation, when both particles are localized at the same position, or in polarization representation when the particles are physically separated (\emph{i.e.}, the overlap between the two wavepackets can be neglected), and can then be treated as distinguishable particles.

\section{Entanglement engineering by discrete-time quantum walks}

\bigskip
In order to illustrate the transversality of the concept of entanglement between identical particles, we engineer a split-step QW protocol for which two indistinguishable photons initially localized around $x=0$ can be measured with a reasonably high probability at different positions after the propagation, and can then be distinguished. With a well choosen set of coin parameters $(\theta_1,\theta_2^{\pm})$, we will show that an initially mode-entangled state leads to  non-vanishing entanglement in polarization between the two photons, whereas an initial state which is separable in terms of modes does not exhibit this behavior. We will refer to this phenomenon by ``entanglement conversion'', from mode entanglement to polarization entanglement, during a temporal transition from indistinguishable to distinguishable particles.

\subsection{Numerical Results}

We numerically compute the split-step QW dynamics of the two following initial states: $|\Psi_A\rangle = |x_1=0, x_2=0\rangle \otimes \frac{1}{\sqrt{2}} \left(|\downarrow \uparrow \rangle +|\uparrow \downarrow \rangle \right)$  and $ |\Psi_B\rangle = |x_1=0, x_2=0\rangle \otimes \frac{1}{\sqrt{2}} \left(|\uparrow \uparrow \rangle + |\downarrow \downarrow \rangle \right) $. The choosen coin parameters are $\theta_1=\frac{\pi}{4}$ and $\theta_2^{\pm}=\pm \frac{\pi}{8}$, and then the topological aspects of the split-step QW are not involved here since the whole system belongs to the same topological zone (see fig. (2)).\\

\begin{figure}[!h]
\begin{center}
\includegraphics[width=0.45\linewidth]{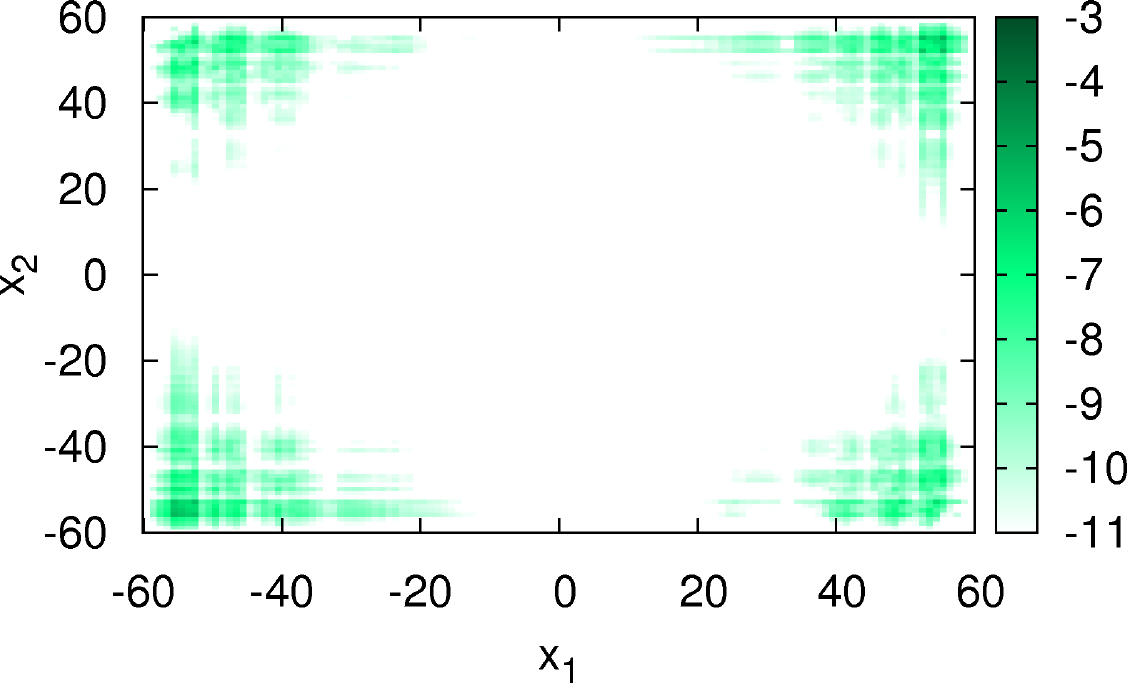} \hspace*{-0.1cm} \includegraphics[width=0.45\linewidth]{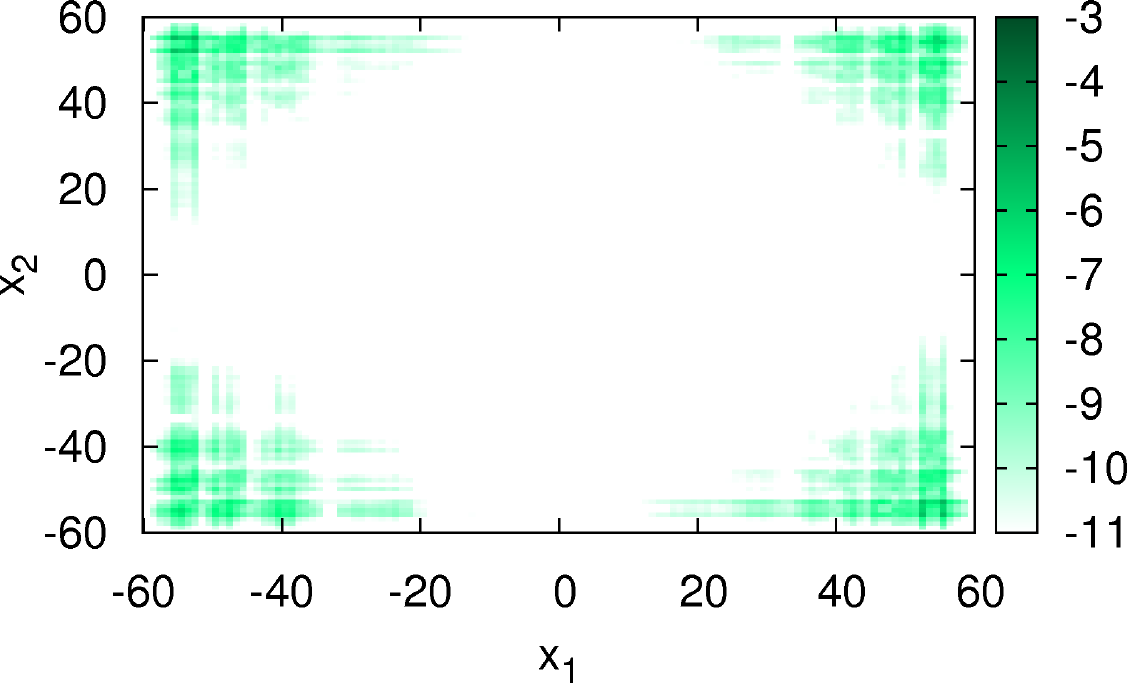}
\caption{Spatial probability distributions in logarithmic scale (color online) of $|\Psi_A\rangle$ (top) and $|\Psi_B\rangle$ (bottom) after 60 iterations.}
\end{center}
\end{figure}

We clearly see on figure (4) that there is almost no difference on the probability distributions between the two cases. Indeed, both present important symmetric peaks of probability on the off-diagonal line : $x_1 + x_2 = 0$, which represents one photon on each side. For each initial state and at each time step $n$, we first project the two-particle state $\left(\hat{U}(\theta_1,\theta_2^{\pm}) \otimes \hat{U}(\theta_1,\theta_2^{\pm})\right)^n |\Psi_{A,B} \rangle$ on the two-position state $| x_{\mathrm{max}},-x_{\mathrm{max}}\rangle$ corresponding to the maximum probability lying on the off-diagonal line, then trace out the position degrees of freedom, and finally normalize the result. This procedure gives a two-polarizations vector $|\psi_0\rangle$, called ``projected" state. Hence, one can build the $4\times4$ ``reduced" density matrix $\hat{\rho}=|\psi_0\rangle\langle\psi_0|$ and evaluate the entanglement between polarizations by computing the polarization Negativity: $\mathrm{N}(\hat{\rho})=\mathrm{Tr}(|\tilde{\rho}|)-1$, where $\tilde{\rho}$ is the partially transposed density operator. \\

\begin{figure}[!h]
\begin{center}
\hspace*{-1cm}
\includegraphics[width=0.65\linewidth]{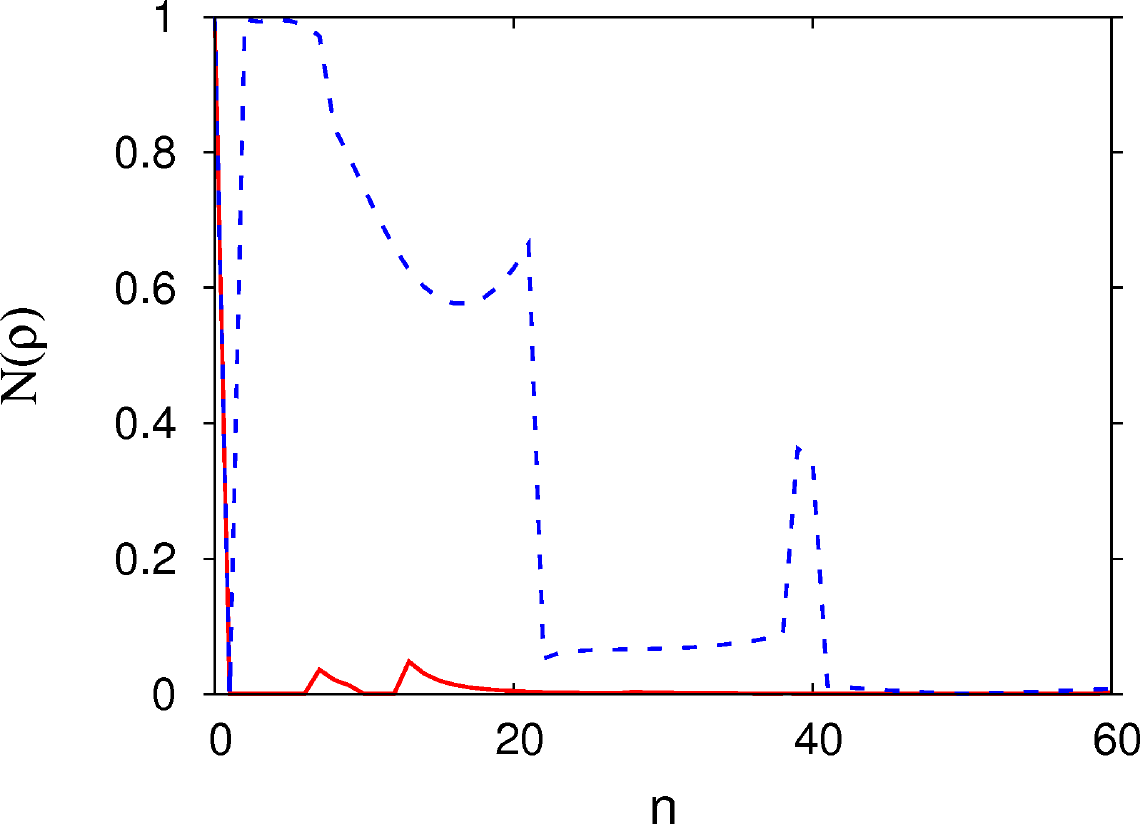}
\caption{The red solid (resp. dashed blue) line represents the computed polarization Negativity of the measured reduced density matrix for initial state $|\Psi_A\rangle$ (resp. $|\Psi_B\rangle$), versus time $n$. In both cases, it is maximum ($\mathrm{N}(\rho)=1$) at the initial time.}
\end{center}
\end{figure}

A radical difference is observed in terms of polarization entanglement during time evolution. In the case of a non mode-entangled initial state ($|\Psi_A\rangle$), some fluctuations of the polarization negativity are observed, close to the minimum value $0$. On the other hand, if the state is initially entangled in modes ($|\Psi_B\rangle$), part of this entanglement is transferred (or converted) to a traditional polarization entanglement during time evolution. 

\subsection{Discussion}

The split-step QW protocol allowed us to propose a simple realization exhibiting a dynamical transition from indistinguishable to distinguishable bipartite system. Additionally, it can offer a nearly direct observation of mode entanglement between identical particles. Let us insist on the fact that at short (\emph{i.e.} experimentally meaningful) times, the measurability of this effect is enhanced because the probability to measure the two photons on the anti-diagonal line is decreasing with time.\\

In the present section, we have avoided the presence of a non-trivial topological structure of the split-step QW in order to focus on the nature of entanglement between identical particles. We will now include this important feature, and focus on the interplay between entanglement and the eventual presence of topological bound states in the spectrum of the single-particle evolution operator.
\bigskip

\section{Entanglement protection by discrete-time quantum walks} 
\bigskip

Let us consider a configuration of ($\theta_1$, $\theta_2^{\pm}$) that crosses one (fig. (6), left panel) or two (fig. (6), right panel) gap closing line(s). 

\begin{figure}[!h]
\begin{center}
\includegraphics[width=0.45\linewidth]{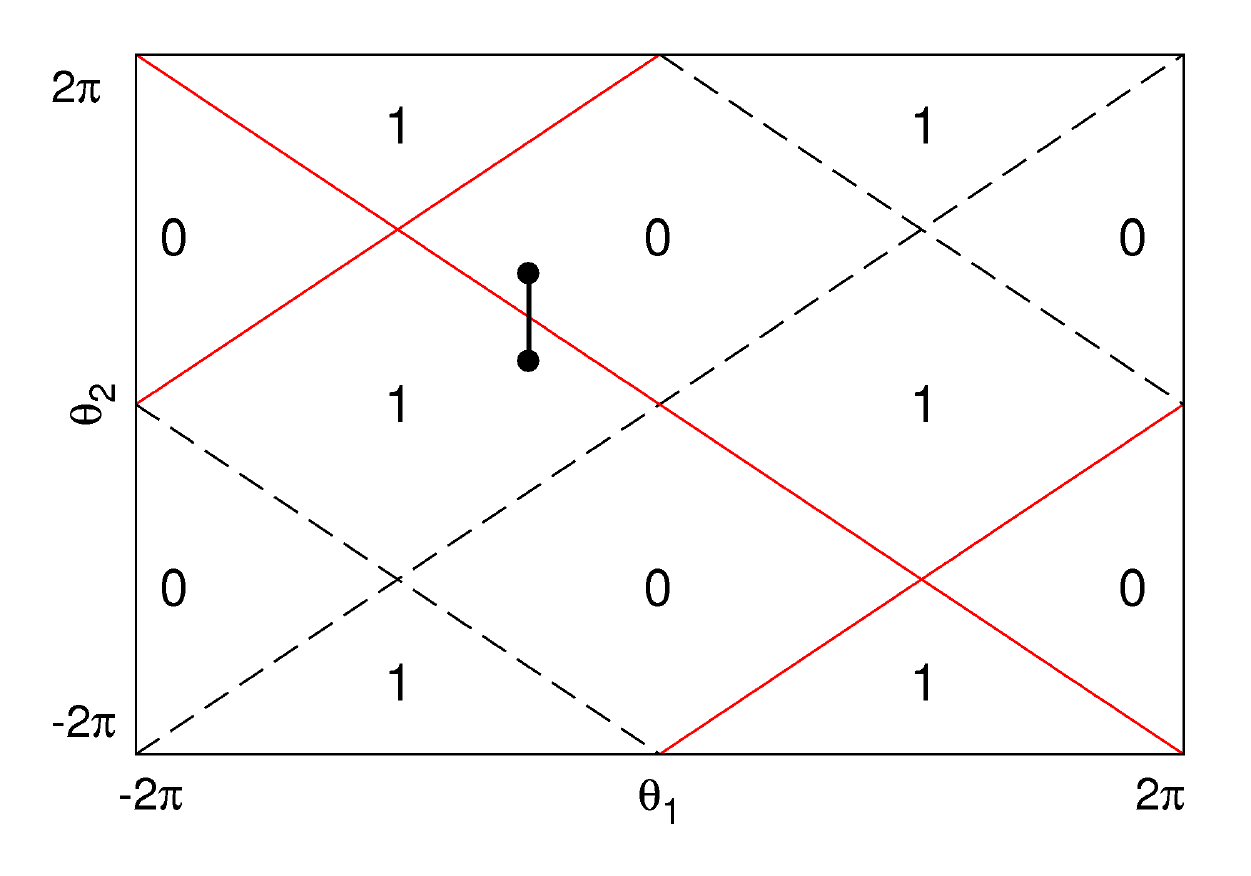} \hspace*{-0.1cm} \includegraphics[width=0.45\linewidth]{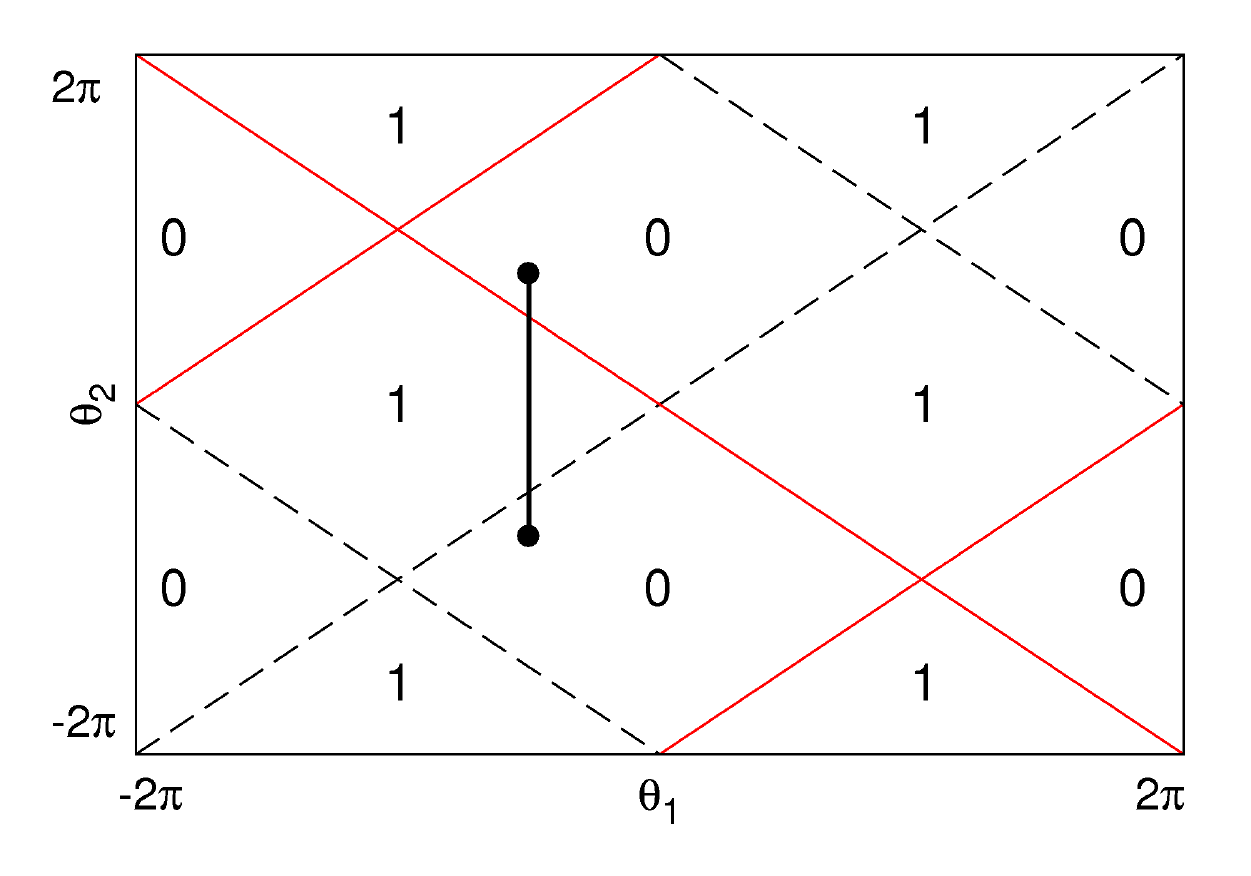}\\
\caption{Configurations of the coin parameters leading to the presence of one (left) or two (right) topological bound states. The red lines describe the trajectories ($\theta_1$, $\theta_2(x)$, $x \in \Re$).}
\end{center}
\end{figure}

Here, we show that when an entangled two-particle state ($|\Psi_B\rangle$) follows a QW protocol for which there is one bound state $|\Psi_{\mathrm{BS}}\rangle $ localized in $x=0$, the two-particle wavepacket presents an important localized part around the origin $x_1=x_2=0$, which can be locally identified to $|\Psi_{\mathrm{BS}}\rangle \otimes |\Psi_{\mathrm{BS}}\rangle$ (see fig. (7), top panel). When two gap closing lines are crossed, the localization phenomenon around the origin is even stronger (see fig. (7), bottom panel). In this case, two different bound states exist in the spectrum of the single particle evolution operator, one $|\Psi^{0}_{\mathrm{BS}}\rangle $ of $0$ quasi-energy, and the other $|\Psi^{\pi}_{\mathrm{BS}}\rangle$ of $\pi$ quasi-energy.

\begin{figure}[!h]
\begin{center}
\includegraphics[width=0.45\linewidth]{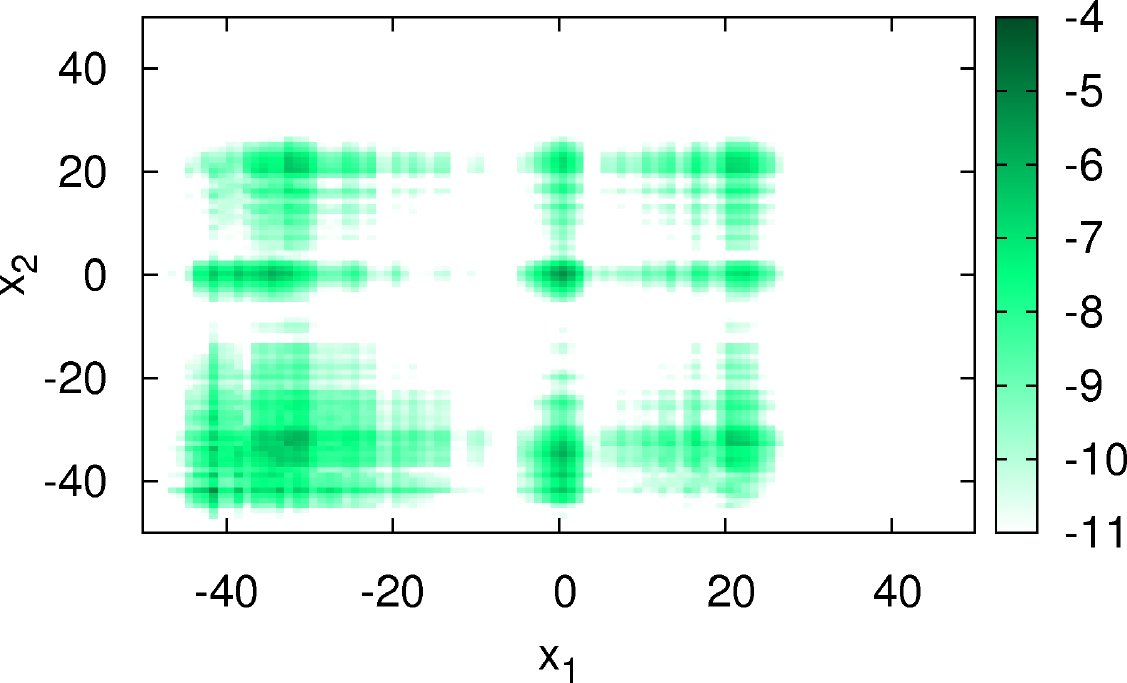}\hspace*{0.1cm}\includegraphics[width=0.45\linewidth]{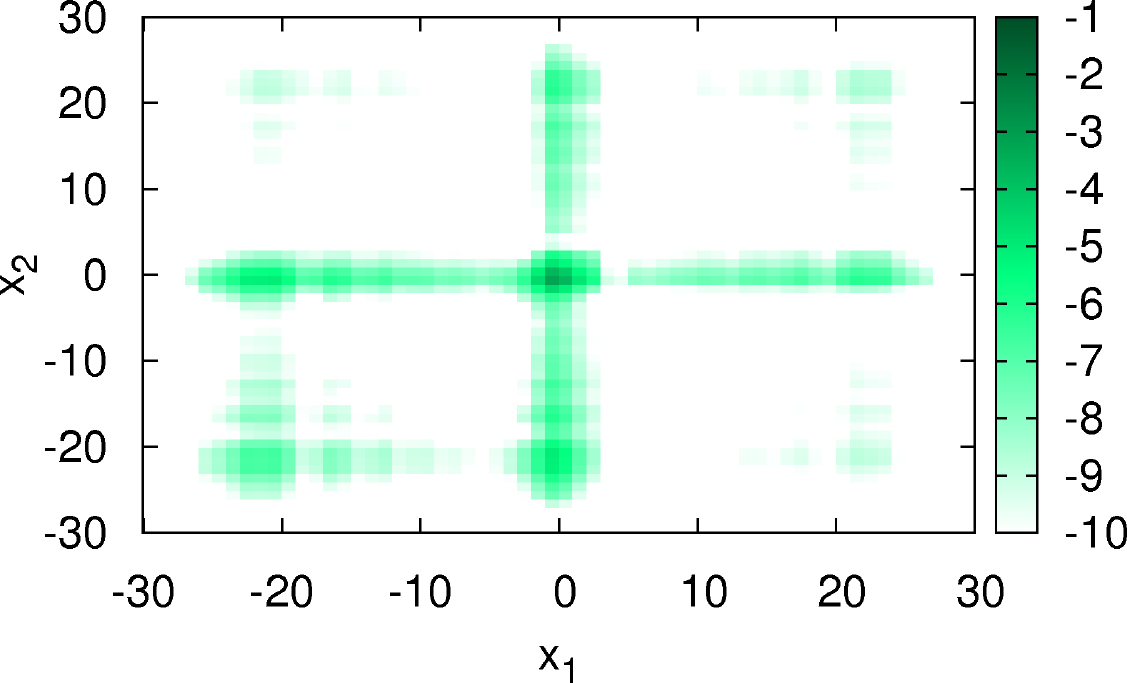}
\caption{Spatial probability distributions in logarithmic scale (color online) of $|\Psi_B\rangle$ after 60 iterations. The spectrum of the single-particle evolution operator contains one single (top) or two (bottom) bound states.}
\end{center}
\end{figure}

We suppose that both photons are now detected at the origin, and compare the mode entanglement during time evolution. This mathematically corresponds to projecting the two particle state on $|x_1=0, x_2=0\rangle$, building up the density matrix represented in modes $\hat{\rho}_\mathrm{m}$, and compute the mode Negativity $\mathrm{N}(\hat{\rho}_\mathrm{m})$.

\begin{figure}[!h]
\begin{center}
\hspace*{-1cm}
\includegraphics[width=0.65\linewidth]{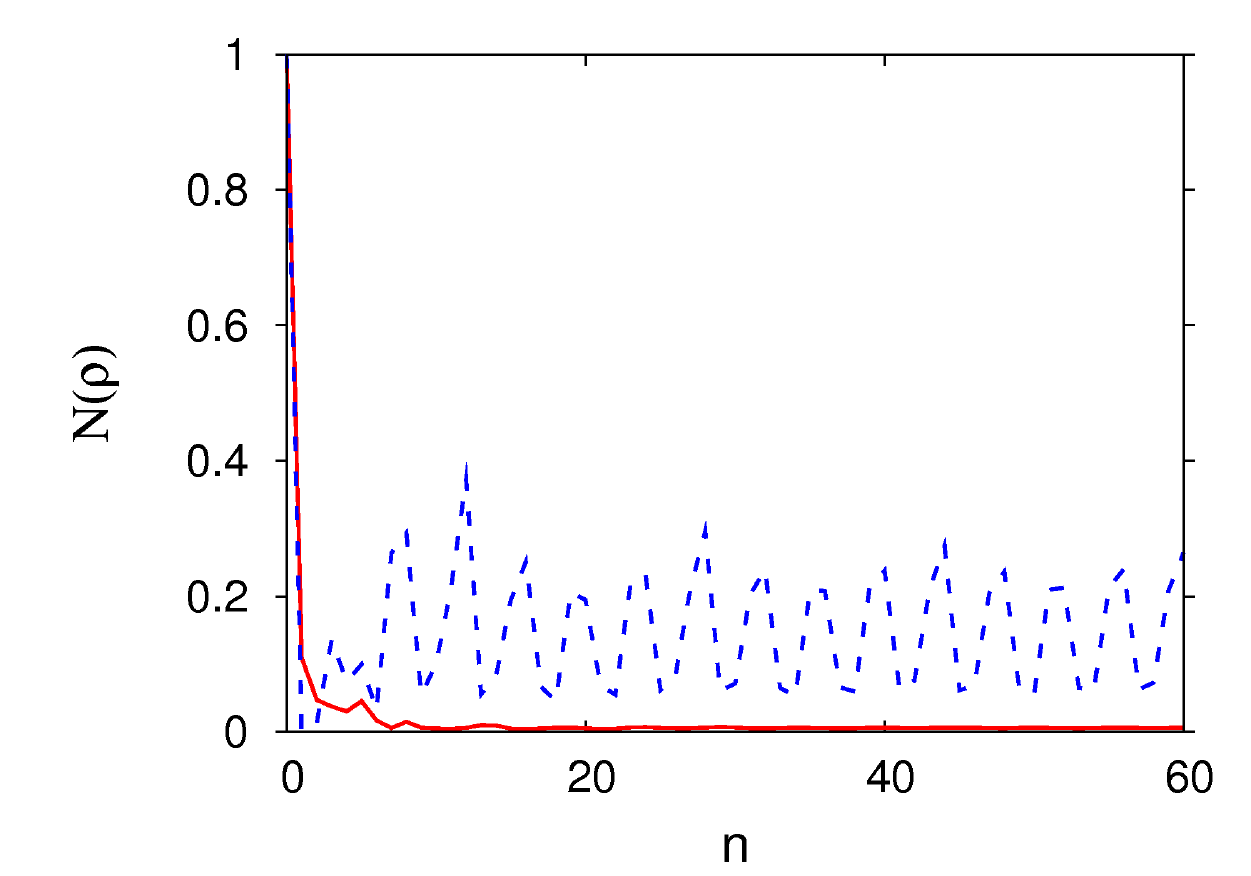}
\caption{Computed Negativity of mode entanglement when the two photons are detected at the origin, for the split-step QW evolution of the initial state $|\Psi_B\rangle$. The red full line corresponds to the presence of a single bound state, and the oscillatory behavior displayed in the blue dashed line is a clear signature of to the presence of two bound states with quasi-energy difference $\pi$.}
\end{center}
\end{figure}

We can easily observe that when two bound states are present in the system, the phenomenon of localization around the origin is clearly enhanced, and thus the probability of detecting both particles in $x=0$ is increased. We can also notice that the entanglement vanishes quickly during time evolution in the first case (1 bound state present), whereas in the other scenario (2 bound states present), the entanglement oscillates between $0.2$ and $0.4$ without decreasing at all. \\

\subsection{Discussion}

\bigskip

Similarly to the single particle case, the main feature of the QW is a ballistic expansion of an initially localized wavepacket. Here, after a certain time, an important part of the wavefunction leaves the origin via ballistic expansion, but what remains in the center is the overlap of the initial state with the bound state(s). In the first case, we can approximate this fraction of the state by a product state: $|\Psi \rangle \approx_{x_1, x_2 \approx 0} |\Psi_\mathrm{BS} \rangle \otimes |\Psi_\mathrm{BS} \rangle$, where $|\Psi_\mathrm{BS} \rangle$ represents the zero energy bound state present in the first case. We can then easily understand the fact that the entanglement rapidly decreases because of the presence of one single bound state.

In the second column case, two bound states are present ($|\Psi^{0}_\mathrm{BS} \rangle$ and $|\Psi^{\pi}_\mathrm{BS} \rangle$), and as previously explained, after a certain time of propagation, the central part of the wavefunction is mainly due to the presence of bound states. When two bound states (or more) are present, we can not conclude that the form of the state is a product: in general, it would be a coherent superposition of the different bound states. This qualitatively explains the important difference observed in the evolutions of entanglement. In the latter case, the non vanishing negativity is directly related to the topological structure of the system and thus, this entanglement should be robust to the presence of impurities, or disorder.\\

The co-existence of pairs of bound states with quasi-energy difference $\pi$ is a robust phenomenon with topological origin, characteristic of periodically driven systems \cite{KitagawaPRB}. In our case we find a similar effect for a static effective Hamiltonian due to the presence of two distinct bound states with quasi-energy $E=0$ and $E=\pi$, when crossing two distinct topological boundaries for our specific choice of coin operators. The co-existence of these bound states, in turn, explains the periodic oscillations  in the Logarithmic Negativity displayed in Figure 8, and represents a novel feature that can be exploited for entanglement protection. The insight into topological protection gained by this approach can help to explore topological properties in  novel correlated materials, where current technology readily enables experimental realizations.
\section{Outlook}

To conclude, in this article we proposed the use quantum walks (QWs) for the controlled engineering and protection of entanglement in bosonic bipartite systems.~The underlying mechanism that allows for manipulation and control of entanglement, as well as for its topological protection, is the strong coupling between internal degrees of freedom in the quantum walker, which plays an analogous role to ``spin-orbit'' coupling in topological systems.~In particular, we showed that QWs can be employed in the transition from bound entanglement to mode entanglement and that such entanglement content can be topologically protected by tailoring the quantum coin, and the initial state for the quantum walker.~The extension of QWs to correlated particles pairs opens the door to simulations of quantum correlations in condensed matter systems with topological order.\\

~The results presented here have direct experimental applications in photonic quantum systems, and could improve the performance of photonic quantum  technologies \cite{hafezi}. In this regard, we note that the phase stability of multi-path QW interferometers is not expected to represent a significant experimental limitation, since topologically protected states are expected to be resilient against local phase noise and perturbations. However, temporal or spatial walk-off in the system with increased number of iterations can play a deleterious dephasing effect which, in combination with losses, can limit the number of iterations. Current bulk optics implementations are limited to $n=7$ iterations at the most \cite{KitagawaNatComm}.    

\bigskip

\section*{Acknowledgments}

G.P. acknowledges financial support from Marie Curie Incoming Fellowship COFUND. S.M. acknowledges Ariel Bendersky and Remigiusz Augusiak for fruitful discussions. This work was supported by ERC grant QUAGATUA, EU integrated project AQUTE, and Spanish MINCIN project FIS2008-00784 (TOQATA).

\bigskip

\section*{References}

\end{document}